\documentclass{article}

\usepackage[nonatbib,preprint]{neurips_2024}

\usepackage[utf8]{inputenc} 
\usepackage[T1]{fontenc}    
\usepackage{hyperref}       
\usepackage{url}            
\usepackage{booktabs}       
\usepackage{amsfonts}       
\usepackage{nicefrac}       
\usepackage{microtype}      
\usepackage{xcolor}         
\usepackage{graphicx}       
\usepackage{algorithm}
\usepackage{algpseudocode}
\usepackage{listings}
\usepackage{graphicx}
\usepackage{caption}
\usepackage{subcaption}
\usepackage{array, xcolor}
\usepackage{wrapfig}
\usepackage{verbatim}
\usepackage{graphicx}
\usepackage{tabularx}
\usepackage{authblk} 
\usepackage{array}
\usepackage{tabularx} 
\usepackage{color}

\title{BladeDISC++: Memory Optimizations Based On Symbolic Shape}

\usepackage{authblk}

\author {Xiulong Yuan$^*$}
\author {Xu Yan$^*$}
\author {Wenting Shen}
\author {Xiafei Qiu}
\author {Ang Wang}
\author {Jie Zhang}
\author {Yong Li}
\author {Wei Lin}
\affil{Alibaba Group}

\begin{document}
\def\thefootnote{*}\footnotetext{Equal Contribution.}
\maketitle

\begin{abstract}

Recent deep learning workloads exhibit dynamic characteristics, leading to the rising adoption of dynamic shape compilers. These compilers can generate efficient kernels for dynamic shape graphs characterized by a fixed graph topology and uncertain tensor shapes. However, memory optimization, although particularly crucial in this large model era, remains relatively underexplored for dynamic shape graphs. The fundamental challenge lies in the lack of precise tensor shapes which are essential in conventional methods such as operation scheduling(op scheduling) and rematerialization. To address this challenge, we propose op scheduling and rematerialization approaches based on symbolic shapes and developed BladeDISC++. Besides, since rematerialization decisions cannot be made solely at compile time when tensor shapes are unknown, BladeDISC++ employs a compilation-runtime combined strategy to optimally address shape dynamics. Evaluations indicate that BladeDISC++ effectively reduces memory usage for dynamic shape graphs, achieving memory consumption comparable to optimizations using precise shapes, thereby promoting the broader adoption of dynamic shape compilers.

\end{abstract}

\section{Introduction}

Dynamic shape compilers are becoming increasingly prevalent due to their ability to optimize deep learning workloads with dynamic characteristics. While systems like TorchInductor\cite{pytorch_dynamic_shape} and Modular\cite{modular_dynamic_shape} have made significant strides in kernel generation, memory optimization still remains underexplored. Conventional methods like op scheduling\cite{meta_olla, op_schedule_for_nas, MAGIS} and rematerialization\cite{checkmate, chen2016trainingdeepnetssublinear, DTR, delta, swap-advisor} (recomputation and offloading included) rely on exact tensor shape to assess the memory impact of ops or rematerialization subgraphs, and make optimization decisions at compile time. However, in the absence of shape values, these methods become unfeasible.

BladeDISC++, built upon a dynamic shape compiler BladeDISC\cite{zheng2023bladedisc}\cite{zhu2021disc}\cite{opensource_disc}, leverages symbolic shapes to tackle the above challenges. With symbolic shapes, BladeDISC++ is able to derive comparative memory impacts of different op sequences and find the optimum scheduling order.
For rematerialization, symbolic shapes are utilized to search for optimum recomputation subgraph at compile time and assist to conduct final rematerialization decisions at runtime.


Our evaluations demonstrate that BladeDISC++ can effectively reduce memory usage for training with dynamic shape graphs compared to BladeDISC. Additionally, BladeDISC++ achieves comparable memory consumption with static shape training while alleviating the overhead of recompilation and tensor padding.

\section{Memory optimizations based on symbolic shapes}

\begin{figure}[htb]
    \centering
    \includegraphics[width=1\linewidth]{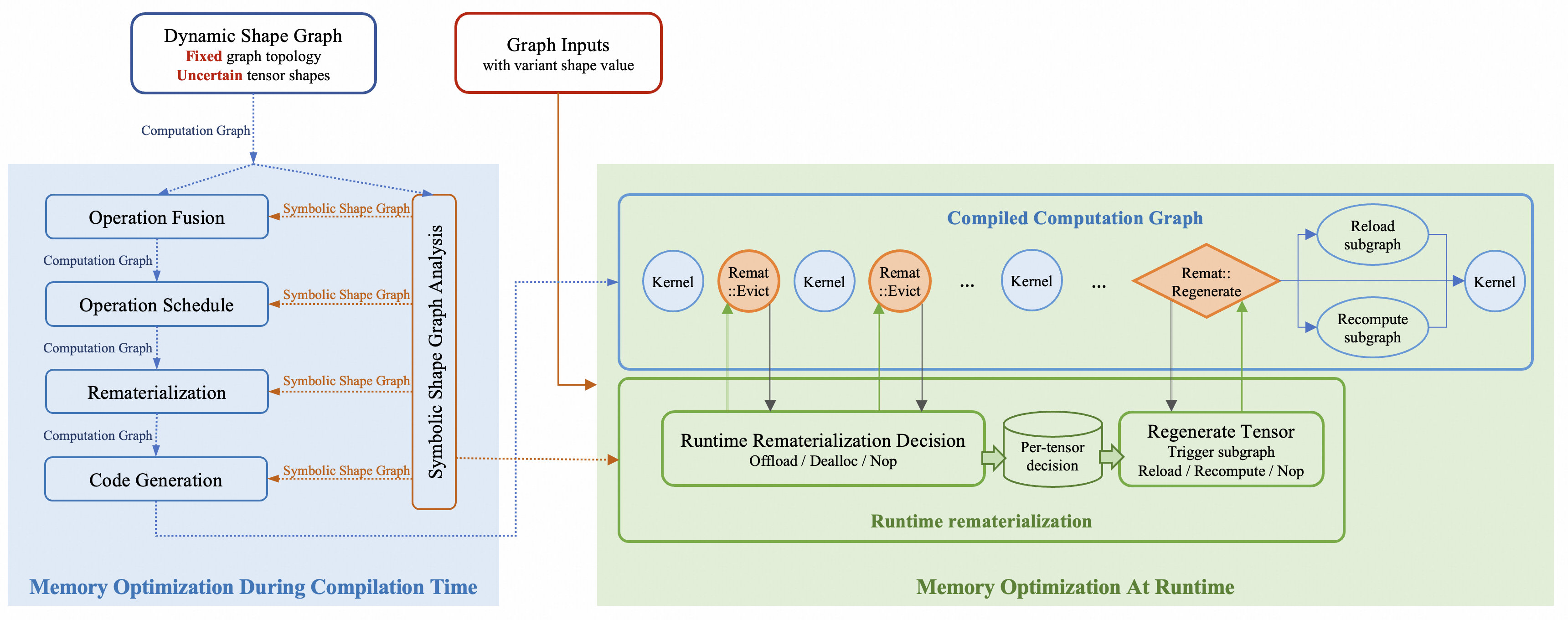}
    \caption{Memory optimizations based on symbolic shapes in BladeDISC++}
    \label{fig:methods_overview}
\end{figure}

As shown in Figure \ref{fig:methods_overview}, given a dynamic shape computation graph, BladeDISC++ first performs symbolic shape analysis to create a global symbolic shape graph that describes the algebraic relationships between shape symbols(in section \ref{methods::dynamic_shape_ir}). Then, the symbolic shape graph, together with the computation graph, undergoes optimization passes including op fusion, op scheduling, and rematerialization for memory optimization. 

As BladeDISC's prior work \cite{zheng2023bladedisc}\cite{zhu2021disc} has tackled the op fusion problem, this paper focuses on op scheduling (in section \ref{methods::operation_schedule}) and rematerialization (in section \ref{methods::auto_rematerialization}).
In particular, with the symbolic shape graph instead of exact tensor shape, BladeDISC++ can still compare memory impacts of different op sequences, and determine whether a recomputation subgraph would benefit memory consumption. Additionally, because a dynamic shape graph might have varying memory footprints across different runs, it is impractical to make rematerialization decisions, such as how much memory to evict, solely at compile time. Therefore, BladeDISC++ explores all rematerialization candidates and searches their corresponding regeneration subgraphs and conduct final rematerialization decisions at runtime. 

\subsection{Symbolic shape graph analysis}
\label{methods::dynamic_shape_ir}


BladeDISC++ systematically analyzes and extracts shape information from the semantics of each op within the dynamic shape computation graph.
It then constructs a global symbolic shape graph to represent the algebraic relationships between shape dimensions through shape value extraction and input-output shape inference.


\definecolor{mygreen}{rgb}{0,0.6,0}
\definecolor{mygray}{rgb}{0.5,0.5,0.5}
\definecolor{mymauve}{rgb}{0.58,0,0.82}

\lstset{ %
  backgroundcolor=\color{white},   
  basicstyle=\footnotesize,        
  breaklines=true,                 
  captionpos=b,                    
  commentstyle=\color{mygray},    
  escapeinside={\%*}{*)},          
  keywordstyle=\color{blue},       
  stringstyle=\color{mymauve},     
}

\lstdefinelanguage{MLIR}{
  keywords={main, @S0, @S1, @C12, @C11008, @C1024, @C4096},
  ndkeywords=
  keywordstyle=\color{blue}\bfseries,
  ndkeywords={Mul, broadcast, dot, reduce, SymbolicDim, dynamic_reshape, export, boolean, throw, implements, import, this},
  ndkeywordstyle=\color{mymauve}\bfseries,
  identifierstyle=\color{black},
  sensitive=false,
  comment=[l]{//},
  morecomment=[s]{/*}{*/},
  commentstyle=\color{mygray}\ttfamily,
  stringstyle=\color{red}\ttfamily,
  morestring=[b]',
  morestring=[b]"
}
\noindent\makebox[\textwidth][c]{%
\begin{minipage}{0.8\linewidth}
\begin{lstlisting}[language=MLIR, basicstyle=\ttfamily\tiny, caption=Example of a dynamic shape graph and its symbolic shape graph, captionpos=b,label=code:symbolicshape, frame=lines]
func.func @main(%arg0: tensor<?,[@S0]>, %arg1: tensor<12x11008>) {
  %1 = broadcast(%arg1) -> tensor<4096x?, [@C4096, @S0]>
  %2 = dynamic_reshape(%arg0, %new_shape) -> tensor<?x12,[@S1, @C12]>
  // The last consumer of %2
  %3 = dot(%2, %arg1) -> tensor<?x11008, [@S1, @C11008]>
  // The last consumer of %3
  %4 = reduce(%3) -> tensor<?, [@S1]>
  %1084 = broadcast(%4) -> tensor<11008x?, [@C11008, @S1]>
  %1085 = broadcast(%arg0) -> tensor<1024x?, [@C1024, @S0]>
}
func.func @symbolic_shape_graph() {
  SymbolicDim @S0
  SymbolicDim @S1
  @S0 = Mul @C12, @S1
}

\end{lstlisting}
\end{minipage}
}

As illustrated in Listing \ref{code:symbolicshape}, BladeDISC++ introduces a \textit{SymbolicDim} op to define a symbolic value, bond to a dimension of a tensor shape in the dynamic shape graph as op attributes, exemplified by \textit{tensor<?x?, [@S0, @S1]>}.
For instance, the equation \textit{@S0 = 12 * @S1} stems from \textit{DynamicReshapeOp} that its input and output tensor have the same number of elements.
 
Comparison between memory sizes of tensors is critical to op scheduling and rematerialization.
BladeDISC++ introduces \textit{SymbolicExpr} to express algebraic representations of symbolic dimensions, allowing for comparative evaluations with a best-effort strategy.
For example, the element number of tensor \textit{\%1084} and \textit{\%1085} can be represented by \textit{SymbolicExpr}s \textit{expr1 = 11008 * @S1} and \textit{expr2 = 1024 * @S0} respectively. As it's already derived from \textit{DynamicReshapeOp} that \textit{@S0 = 12 * @S1}, \textit{exp1} can be simplified to \textit{132096 * @S0}, thus BladeDISC++ can infer that \textit{expr1} is less than \textit{expr2}.

\subsection{Operation scheduling}
\label{methods::operation_schedule}
\begin{wrapfigure}{r}{0.50\textwidth} 
    \centering
    \includegraphics[width=1\linewidth]{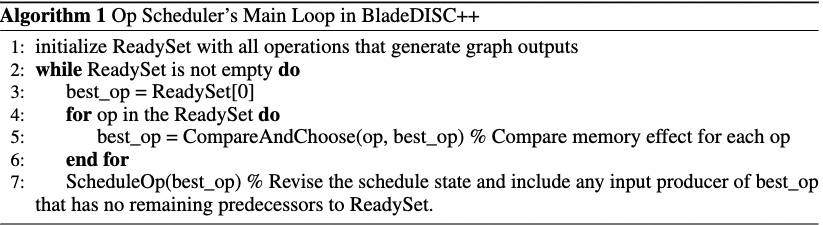}
    \caption{OpScheduler algorithm main loop}
    \label{fig:uni_scheduler}
\end{wrapfigure}
Op scheduling tries to find a memory-efficient op sequence from the original computation graph. Existing scheduling algorithms\cite{XLA} often traverse the computation graph and select an op from a \textit{ReadySet} (including ops whose predecessors have already been scheduled) at each step. The selection is mainly based on comparing different ops' memory impact, which is determined by the difference between bytes freed and allocated after scheduling a specific op. BladeDISC++ adopts a similar methodology, emphasizing the computation and comparison of memory impact among different ops with the absence of exact tensor shapes in dynamic shape graphs. Specifically, in BladeDISC++, the memory impact for each op is calculated using symbolic shapes and thus expressed as a \textit{SymbolicExpr}. These \textit{SymbolicExpr}s are then compared to each other with the help of symbolic shape graph.

In Listing \ref{code:symbolicshape}, for example, the \textit{DynamicReshapeOp} and \textit{DotOp} appear in the \textit{ReadySet} at a specific step. \textit{DotOp}, as the last consumer of \%2 and producer of \%3, has a memory impact of \textit{10996 * @S1} due to the allocation for \%3 and deallocation for \%2. The \textit{DynamicReshapeOp}'s memory impact, on the other hand, is \textit{4096 * @S0} because scheduling it only involves allocation for \%1. 
To compare two \textit{SymbolicExpr}s containing different sets of symbols, we first simplify the \textit{SymbolicExpr} of \textit{DynamicReshapeOp}'s memory impact based on \textit{@S1}
to \textit{49152 * @S1} with the same procedure described in \ref{methods::dynamic_shape_ir}, then it can be determined that the \textit{DynamicReshapeOp} has a higher memory impact than the \textit{DotOp}. 

When it's unfeasible to compare two memory impact \textit{SymbolicExpr}s, we resort to a commonly used strategy: selecting the op that results in smaller overall tensor lifetimes based on the graph topology.

\subsection{Rematerialization}
\label{methods::auto_rematerialization}
Conventional rematerialization methods\cite{checkmate, delta, chen2016trainingdeepnetssublinear}involve algorithms to determine which tensors to be evicted earlier to alleviate memory pressure, as well as how to perform subsequent regeneration, either through reloading or recomputation. These methods also include a search process to identify optimal recomputation subgraphs by evaluating their memory impacts. Notably, tensor rematerialization may negatively affect end-to-end performance, so it should only be employed when the graph's execution risks exceeding the memory limit.  However, a dynamic shape graph, with undetermined tensor shapes, can exhibit varying peak memory usage across different runs. Some runs may not need rematerialization since they remain within memory limits, while others may need. It's impractical to make all decisions solely during compilation. Furthermore, the lack of exact shapes raises challenges in assessing the memory impacts of potential recomputation subgraphs.

To address these issues, BladeDISC++ utilizes a combined compilation-runtime strategy based on symbolic shapes to best manage shape dynamics across graph runs. During compilation, it explores all potential rematerialization candidates and identifies their corresponding regeneration subgraphs, which are then inserted into the original computation graph as different execution branches. Final decisions regarding which tensor to evict and the associated regeneration method are made at runtime.

During compile time, as illustrated in Figure~\ref{fig:methods_overview}, BladeDISC++ inserts a \textit{Remat::EvictOp} after each op, checking if any active tensors at that point need to be evicted to alleviate memory pressure. For each candidate tensor, regeneration subgraphs, including those for reload and recomputation, are also generated. While reloading only involves a host-to-device (H2D) instruction and is memory-neutral, searching for recomputation subgraphs requires careful evaluation since sub-optimal choices may even increase peak memory usage. BladeDISC++ uses a standard search process but assesses memory impact of potential subgraphs based on \textit{SymbolicExpr}.

Taking recomputation subgraph searching for \textit{\%4} in Listing \ref{code:symbolicshape} as an example. Starting from the \textit{ReduceOp}, BladeDISC++ determines the memory impacts: \textit{-11007 * @S1} for just the \textit{ReduceOp}, \textit{-11 * @S1} with the addition of the \textit{DotOp}, and \textit{1 * @S1} when the \textit{DynamicReshapeOp} is included. Although exact shape values are unknown, BladeDISC++ can still ascertain that the last recomputation subgraph is memory-efficient, whereas the others are not.

Then, BladeDISC++ inserts \textit{Remat::RegenerateOp}s, along with the corresponding regeneration subgraphs (both reload and recompute), before each candidate tensor's subsequent consumers. The \textit{Remat::RegenerateOp} checks whether a candidate tensor is evicted and its regeneration method,

At runtime, BladeDISC++ monitors memory usage throughout kernel execution. Each time an \textit{EvictOp} is triggered, BladeDISC++ checks the current memory usage and performs an on-the-fly analysis of all candidate tensors provided by the \textit{EvictOp} when the memory limit is about to be surpassed. The final decisions on which tensor from the above candidates needs to be evicted as well as the corresponding regeneration method are made by considering factors such as memory savings and end-to-end performance impact, following a similar approach as outlined in \cite{delta}. Subsequent \textit{Remat::RegenerateOp}s then query these decisions and determine which regeneration subgraphs need to be triggered. 

\section{Evaluation}
For our evaluation, we conducted experiments on the supervised fine-tuning of Llama-2-1b, a tailored model from the official Llama-2-7b\cite{Llama-2-7b} with the only change that decreasing \textit{num\_hidden\_layers} from 32 to 4, on an Alibaba Cloud ecs.gn7-c12g1.3xlarge instance\cite{test-bed}(with 40GB GPU RAM) using the CodeAlpaca-20K dataset \cite{CodeAlpaca-20k}. CodeAlpaca-20K contains samples with text lengths ranging from approximately 100 to 3000 characters. In each training iteration, a fixed number of randomly selected samples are assembled into a batch, resulting in variable batch shapes across different iterations.

To assess the effectiveness of BladeDISC++, we compared memory usage and end-to-end performance in dynamic shape training using BladeDISC++ against both dynamic and static shape training using BladeDISC. For static shape training, following common practice, input sequences are padded to nearest power of 2 in length to balance redundant computation and compilation overhead. Besides, in our experiments, we deliberately set the largest bucket size equal to the longest sequence length in the dataset to investigate whether we can achieve comparable memory optimization results using symbolic shapes instead of exact shapes. 

The experimental results indicate that BladeDISC++ can effectively reduce peak memory consumption for dynamic shape training. Furthermore, BladeDISC++ demonstrates comparable memory consumption to static shape training while also improving end-to-end performance by alleviating the overhead of recompilation and input bucketing.

\begin{table}[htbp]
\centering
\caption{Training throughput of Llama-2-1b on CodeAlpaca-20K(tokens/second)}
\begin{tabular}{|c|c|c|c|}
\hline
 Batchsize & 14 & 16 & 18 \\ \hline
\small{BladeDISC(dynamic shape training)}& \small{5662.34(38.20 GiB)}& \textcolor{red}{OOM} & \textcolor{red}{OOM}  \\ \hline
\small{BladeDISC(static shape training)} & \small{5242.02(35.75 GiB)} & \small{5429.38(37.71 GiB)} & \small{5103.31(38.92 GiB)}  \\ \hline
BladeDISC++ & \small{5749.20(35.76 GiB)} & \small{6078.71(37.89 GiB)} & \small{5738.79(39.18 GiB)} \\ \hline
\end{tabular}
\label{tab:e2e_result}
\end{table}

\section{Conclusion}

This paper shares our industry experience in optimizing memory for dynamic shape graphs . We proposed op scheduling and rematerialization based on symbolic shapes and developed BladeDISC++. Evaluations show that BladeDISC++ can effectively reduce memory usage for dynamic shape training and can achieve comparable memory optimization results to static shape training. As far as we know, this is a pioneering effort in this area, and we aspire that it will support the compiler community in managing dynamic shape workloads and promote wider use of dynamic shape compilers.

{
\small
\bibliographystyle{plain}
\bibliography{references}
}

\appendix


\end{document}